\documentclass[aps,twocolumn,
showpacs,groupedaddress,floatfix]{revtex4}
\usepackage[dvips]{epsfig}

\begin{document}

\title{Structure prediction based on {\it ab initio} 
simulated annealing for boron nitride}

\author{K. Doll, J. C. Sch\"on, M. Jansen}
\affiliation{Max Planck Institute for Solid State Research, Heisenbergstr. 1, 
D-70569 Stuttgart, Germany}

\date{\today}

\begin{abstract}
Possible crystalline modifications of chemical compounds at low temperatures
correspond to local minima of the energy landscape. Determining these minima
via simulated annealing is one method for the prediction of crystal
structures, where the number of atoms per unit cell
is the only information used. It
is demonstrated that this method can be applied to covalent systems, at
the example of boron nitride,
using {\it ab initio} energies in all stages of the optimization, i.e.
both during the global search and the subsequent local optimization. Ten low
lying structure candidates are presented, including
both layered structures and
3d-network structures such as the wurtzite and zincblende types, as well
as a structure corresponding to the $\beta$-BeO type.
\end{abstract}



\maketitle

\section{Introduction}

The knowledge of the
crystal structure of a solid compound
is one of the basic questions in solid state
theory\cite{Maddox88,Cohen89,Catlow90,Schoen96b,Jansen02b}.
Since the early 1990's, an effort has been made to develop methods to
predict structures of solids
without any experimental information about the structure.
The starting point is the realization that any (meta)stable modification
of a (solid) compound corresponds to a locally ergodic region on the
energy landscape of the chemical system. For low temperatures, such
regions are centered on local minima of the energy that are surrounded
by sufficiently high energy barriers. Since both the thermodynamically
stable and the multitude of kinetically stable modifications are of interest,
the global search is not restricted to the determination of the global
minimum, but also includes local minima \cite{Schoen96b}.

The most common methods used for the structure prediction of solids
are simulated annealing
\cite{Kirk83,Czerny85,Pannetier90},
genetic algorithms \cite{Holland75,Johnston04,Oganov06,Woodley07},
basin hopping \cite{Nayeem1991,Wales1997}, or the recently introduced
metadynamics \cite{Laio2002}.

Structure prediction usually involves a huge amount of CPU time, and therefore
efficient ways to keep the calculations tractable have to be found.
Thus, the procedures were initially split in two steps: first,
a global search on the potential surface was performed. The energy was
evaluated with empirical potentials, e.g. Coulomb and Lennard-Jones
potentials, or chemically/physically motivated cost functions. 
After the global search, e.g. using simulated annealing,
possible structure candidates
were locally optimized on the {\it ab initio} level, usually with
density functional theory. 

Empirical potentials are very efficient, but also have various
drawbacks: they work reasonably well for ionic systems, but less
for covalent systems; and some knowledge of the expected bond type is 
required in
advance to choose the potential. Recently \cite{DollPCCP}, 
we demonstrated that
a full {\it ab initio} treatment is feasible in both stages, i.e. the global
search and the subsequent local optimization can both be performed on
the {\it ab initio} level. The system considered (lithium fluoride) 
was chosen for several
reasons: the small number of electrons leads to fast calculations,
and the ionicity of the system makes convergence easy. Moreover,
the system had earlier been studied with model potentials
\cite{Schoen95,Cancarevic08}, and it turned out that the relevant minima
were the same when full {\it ab initio} structure prediction was performed
\cite{DollPCCP} (for a brief summary see also \cite{Dollfuerpisani}).

In the present work, boron nitride (BN)
was chosen as an example for a covalent system.
This is a significant extension of the previous work, since a covalent system
is much more difficult to study: covalent bonds between the atoms have
to be formed, and convergence for random structures is much more difficult
in this case. At zero pressure, BN has a hexagonal structure, with
space group 194, see e.g. \cite{Pease1952}. Under pressure, it may
transform to a zincblende \cite{Wentorf1957} 
or a wurtzite structure \cite{Bundy1963}, and a corresponding
phase diagram was obtained \cite{CorriganBundy1975}. There is, however,
an ongoing discussion concerning the correct phase diagram, 
see e.g. \cite{Mujica2003}. 
Also, a rhombohedral structure was found
(a layered structure similar to the hexagonal structure)
\cite{Ishii1981},
and turbostratic boron nitride with a random arrangement of the layers 
has been reported \cite{Thomas1963}. These structures
are summarized in table \ref{expstructures}. For overviews, see
also \cite{MSF,StructureBonding102}.
Early ab-initio calculations were performed from the mid 1980's onwards, 
e.g.
\cite{Wentzcovitch1986,Park1987,Catellani1987,Orlando1990,Xu1993,Christensen1994,Furthmueller1994,Albe1997,Kern1999}. 

The present task is for several reasons far from straightforward:
first, the CPU time has to be reduced by a large factor. As we showed
for the system LiF \cite{DollPCCP}, 
simply performing simulated annealing with standard
{\it ab initio} calculations would lead to CPU times of the order of 2 years
for a single run, and often hundreds of such runs need to be
performed. Therefore,
very subtle methods have to be employed to reduce the CPU time
for the {\em ab initio} calculations, without losing too much accuracy.
In our earlier study of LiF, 
very careful tests were required in order to
reduce the CPU time to a few days. Similarly, many tests have to be
performed to reduce the required CPU time in the case of BN to reasonable
values. Secondly, one needs a strategy to converge the system in all steps
of the global exploration: at the beginning of the search,
the unit cell has a very large volume,
and the atoms are at random positions within the cell, 
i.e. the configuration is like in a gaseous state.
The total energy calculation at such a geometry has to be converged,
and it has to be converged for all the subsequent steps of the simulated
annealing procedure. This has to be done in an automatic way, since thousands
of calculations at successive geometries are performed.
Finally, it is a very interesting and important question whether
this procedure will find all the very different
structure types: the layered structures,
and the three-dimensional structures such as wurtzite and zincblende.

\begin{table}
\begin{center}
\caption{\label{expstructures}The experimentally found structures.}
\vspace{5mm}
\begin{tabular}{cccccc}
\hline\hline
 structure type &  \multicolumn{3}{c}{cell parameters and fractional
coordinates} 
\\ \hline
hexagonal BN \cite{Pease1952} & a=2.504 \AA, c=6.661 \AA \\
           & B (1/3, 2/3, 1/4) \\
           & N (1/3, 2/3, 3/4) \\ \\
zincblende \cite{Wentorf1957} & a=3.615 \AA \\
           & B (0, 0, 0) \\
    & N (1/4, 1/4, 1/4)  \\ \\
wurtzite \cite{Bundy1963} & a=2.55 \AA, c=4.20 \AA \\ \\
rhombohedral \cite{Ishii1981} & a=2.52 \AA, c=10.02 \AA \\
\hline\hline
\end{tabular}           
\end{center}
\end{table}

\section{Method}
\label{methodsection}

The general method consists of several steps:
first, a simulated annealing run with a subsequent stochastic quench
is performed, to identify possible
candidate structures. This is followed by a local optimization based
on analytical gradients. Finally, the symmetry and space group are identified.
This is repeated many times, in order to identify as large as possible
a set of
structure candidates, and to obtain some statistics about the structures
found.

The details for the BN calculations are described in the following
paragraphs.
4 boron and 4 nitrogen atoms were placed at random positions
in a large unit cell which
was initially cubic with the cell parameter a=6.23 \AA. 
This initial volume is computed by first employing the atomic/ionic 
radii to estimate the total volume occupied by the atoms in the solid state
and then multiplying this number by a sufficiently large factor
(typically between 3 and 10), to allow the atoms enough freedom
to reach any arrangement independent of the initial random placement
in the simulation cell. This factor was varied in preliminary calculations, 
where it was found that choosing a larger volume than the one selected 
(6.23$^3$ \AA$^3$ = 242 \AA$^3$, which is already by a factor of $\sim$5 larger
than the volume of e.g. the zincblende structure) did not lead to
significantly different results.

Each simulated annealing run had a length of 12500 steps, and the
initial temperature of 1.00 eV (corresponding to 11604 Kelvin) 
was reduced to $\sim$ 0.78 eV at the end of the run. 
The length is
thus somewhat longer than in the case of LiF \cite{DollPCCP},
because it
is more difficult to approach possible candidate structures in 
a system with covalent bonds. 
The simulated annealing was
followed by a quench with 5000 steps, i.e. a simulated annealing run
with a temperature of 0 eV, which means that only downhill moves are allowed
during the quench. The quench thus moves the geometry 
obtained after the simulated annealing further towards a local minimum.
The moves were chosen as: moving individual atoms (70\%), exchanging
atoms (10\%), changing the lattice parameters with fixed fractional
coordinates (10\%), changing the lattice parameters with fixed
cartesian coordinates (5\%), and changing the origin (5\%, this
move is important if subsequently the cell parameters change).
No symmetry was prescribed during the simulated annealing and quench runs,
i.e. the space group was always $P1$.

A minimum distance between two atoms (given by
the sum of the radii of the atoms, multiplied
with 0.7) was prescribed in order
to avoid unrealistic geometries which may lead to
numerical
instabilities. The radii used were based on tabulated values
for atomic and ionic radii, as a function of charge, and the 
Mulliken charge computed for the previous configuration.
In those moves which change the lattice constant,
the probability of reducing the lattice constant was enlarged to 60\%,
to speed up the reduction of the cell size.

The {\it ab initio} 
calculations were performed with the CRYSTAL06 code \cite{Manual06},
which is based on local Gaussian type orbitals.
Two basis sets were used during the simulated annealing runs, starting from 
a $[3s2p]$ basis set for B and N, with the inner $[2s1p]$ exponents  as
in \cite{Binkley1980}. In the case of basis set I, additional $sp$ exponents
of 0.4 for B and 0.3 for N were chosen, and the outermost exponent
of the B 2$sp$ contraction (0.4652) was removed.
In the case of basis set II, the outer $sp$ exponents
were chosen as 0.25 for B and 0.297 for N. The basis sets are given
in table \ref{Basissetssimannlocopt}. In the stage of the local optimization,
the basis sets used were the
$[3s2p1d]$ basis sets from
\cite{Orlando1990} (basis set III in table \ref{Basissetssimannlocopt}).

The basis sets during the simulated
annealing are therefore
chosen slightly different from the ones used in the local
optimization (less diffuse functions, no
polarization functions), to speed up the global search
which is the time-consuming part of the procedure. 
To test these basis sets, 
as a preliminary step, the energies of the wurtzite and of the
layered $B_k$ structure were computed with various basis sets
(table \ref{HFbasissimann}). With
basis set III, which is used during the local optimization, the
$B_k$ structure is more favorable by $\sim$ 30 $mE_h$, i.e. $\sim$ 0.8 eV
(1$E_h$ = 1 hartree = 27.2114 eV).
The smallest basis set I gives preference to the wurtzite
structure instead, by 33 $mE_h$ (0.9 eV), and basis set II gives preference
to the $B_k$ structure, by 8 $mE_h$ (0.2 eV). One might thus fear
that basis set I would not yield the layered structures during the global
search; however, this is
not the case, as will be shown in the results section; 
and there does not seem to be a strong bias due to the basis set.
Basis set I is however advantageous because
calculations are roughly twice as fast as with basis set II.

\begin{table}
\begin{center}
\caption{\label{HFbasissimann}
The Hartree-Fock energies (in hartree, per four formula units)
of the $B_k$ and wurtzite structures,
computed with the small basis sets 
used during the global optimization (basis set I, II),
in comparison with the energy obtained with the
basis set used for the local optimization 
(basis set III). The geometry was fixed at the computed equilibrium
geometry of basis set III (table \ref{Structuresfound}).}
\vspace{5mm}
\begin{tabular}{ccc} \hline\hline
structure & basis set & total energy \\ \hline
$B_k$ &  I  & -316.6687 \\
       &  II & -316.7495 \\
       &  III & -316.8753  \\
wurtzite & I & -316.7014 \\
       &   II & -316.7419 \\
        &  III & -316.8452 \\ \hline\hline
\end{tabular}
\end{center}
\end{table}

Concerning the choice of the {\it ab initio} 
method, it has to be taken into account that
convergence at a random geometry is necessary during the global exploration
stage. As mentioned earlier, the initial geometry has a cell volume
$\sim$ 5 times larger than the experimental volume, and the atoms
are randomly arranged. A typical band structure of such
a geometry is very localized due to the large interatomic distances,
and completely different from the band structure of the experimental
geometry in the solid state. Therefore, convergence is absolutely non-trivial.
It turned out that
convergence was best achieved with the Hartree-Fock approach, due
to the fact that the band gaps are usually very large.
Indeed, the band structure and corresponding densities of
states display band gaps of $\sim$ 6 eV (Hartree-Fock), $\sim$ 0.5 eV (B3LYP),
and $\sim$ 0.1 eV (LDA) for this {\em initial} structure. Note that
this gap corresponds to a random initial structure and
is very different from the gap of the final structure; but
it is necessary to converge a calculation for this random initial geometry,
and for all the geometries subsequently generated, until the end of
the simulated annealing and quench.
For comparison, calculations with the hybrid functional
B3LYP were found to be
much more difficult to converge, and a large mixing ratio was required:
90\%, in 
combination with the Anderson mixing scheme; 35\% was sufficient in
the case of Hartree-Fock (the mixing ratio is 
the ratio with which the previous Fock operator is added to the
new one, in order to achieve convergence).
This leads to many iterations and thus a large CPU time. The local
density approximation (LDA) was very difficult
to converge for random atom arrangements, and needed more $\vec k$ points
and level shifting. 

Interestingly, even for the initial structure consisting of widely spaced,
nearly isolated atoms, it 
was sufficient to use the restricted Hartree-Fock approach, 
i.e. it was not necessary to
take into account spin-polarization which would be the case for a free
atom.

The thresholds for integral selection were enlarged from 10$^{-6}$, 10$^{-6}$,
10$^{-6}$, 10$^{-6}$, 10$^{-12}$ to 10$^{-4}$, 10$^{-4}$, 10$^{-4}$,
10$^{-4}$, 10$^{-8}$, respectively,
and the self consistent field cycles were stopped 
when the difference between two subsequent cycles was below  10$^{-4}$
$E_h$. A mesh with 4$\times$4$\times$4  $\vec k$-points was used.
The error associated with the $\vec k$ mesh can be estimated by
computing the energy difference when changing the lattice constant: 
e.g. changing the lattice constant of BN in the zincblende structure
from 3.7 to 3.6 \AA \ changes the energy by 0.01368 $E_h$ for four
formula units with a 4$\times$4$\times$4  mesh, and by
0.01436 $E_h$ with a 8$\times$8$\times$8 mesh. The associated error with
the mesh is thus 0.01436-0.01368=0.00068 $E_h$ and reasonably small.

The simulated annealing and subsequent quench 
are the time consuming parts, and a
single run takes of the order of one week on a single CPU. The same
approach would have been feasible with the B3LYP functional, but at
a much higher cost, for the reasons discussed above
(around a month instead of one week CPU time). It
appeared therefore more reasonable to perform four times as many
runs (here: around 329, see table \ref{StructuresEnergies}) 
using the Hartree-Fock approach, 
as with the B3LYP approach where around 80 runs
would have been feasible with a comparable total CPU time.

The local optimization is not very time consuming and
was done with default parameters for the integral
selection and the self consistent field cycles. The full geometry
optimization can by now be routinely performed with analytical gradients
\cite{IJQC,CPC,KlausDovesiRO,KlausDovesiRO1d2d,Mimmo2001}
as implemented in the CRYSTAL06 release. The local optimization,
starting from the structure after the quench, was done both
at the HF level and at the LDA level; in nearly all the cases, the
resulting final minimum structures turned out to be the same. In addition,
for these final structures, also a B3LYP optimization
was performed, in order to compare Hartree-Fock, B3LYP and LDA.
The basis set used was basis set III in
table \ref{Basissetssimannlocopt}. In addition,
in appendix \ref{basissetextrapolation}, larger basis sets were tested,
for comparison.

The symmetry was analyzed with the program KPLOT \cite{KPLOT}
where algorithms to find the symmetry and space group
\cite{Hundt99a,Hannemann98a} are implemented.

For the most important structures,
the enthalpy was computed, in order to investigate the pressure dependence of
the phases.

\section{Results and discussion}

The most relevant structures found are displayed in figures
\ref{schichten},\ref{186-216-136},\ref{62-8}, and \ref{14-9}, visualized with
XCrysDen \cite{XCrysDen}. 
Optimized geometries are given in table \ref{Structuresfound}.
First, the experimentally observed $B_k$ structure was obtained (so-called
hexagonal boron nitride, space 
group 194). Closely related are two additional layered structures,
with space group 160 and 187, respectively. 
In the $B_k$ structure, sheets are made
of edge-connected six-membered rings of 3 boron and 3 nitrogen atoms 
in alternating sequence (see figure \ref{schichten}, left).
The neighboring sheets
are stacked vertically below and above, with alternating atoms
(i.e. N sits vertically above B, and vice versa; the stacking order is ABAB). 
In space group 160,
the same sheets are formed, but only three atoms have neighbors in the
layer above, and the other three atoms have neighbors in the layer below
(the stacking order is ABCABC, i.e. like in rhombohedral BN
\cite{Ishii1981}). 
The structure with space group 187 has
stacking order ABAB, where again three 
atoms have neighbors vertically above and
below in the next layer. These three layered
structures have a very similar total energy,
and also the enthalpy as a function of pressure looks very similar
(see figure \ref{LDAenthalpies}).

The wurtzite structure is displayed in figure \ref{186-216-136}, left.
The zincblende structure is displayed in figure \ref{186-216-136}, middle,
and has
a similar energy as the wurtzite structure.
At zero pressure, 
the energies of the wurtzite and zincblende structure
are comparable to that of the layered structures in figure
\ref{schichten}.

The structure with space group 136 has six-membered rings (3 B, 3 N,
alternating), but
also rings with 4 (2 B, 2 N) atoms. 
This leads to angles close to 90$^\circ$ and
a less favorable energy (see figure \ref{186-216-136}, right). 
This structure corresponds to the $\beta$-BeO type \cite{Smith1965}, 
which demonstrates
that the method presented gives reasonable low-lying structure candidates:
as Be has one electron less than B, and O one more than N, it makes
sense that such the $\beta$-BeO structures is found as a
candidate structure also for BN. Under ambient conditions,
BeO crystallizes in the wurtzite structure,
and the $\beta$-BeO structure is found as a high temperature phase
\cite{Smith1965}.

The structure with space group 62 has four-, six- and eight membered
rings. The topology is similar to the one of the aluminum
network in the SrAl$_2$ structure under ambient pressure
(space group 74, Imma): one has to replace one aluminum atom with boron,
the neighboring one with nitrogen, and discard the strontium.
This is reasonable, as the two aluminum atoms
obtain two electrons from strontium and thus have 8 valence electrons together,
i.e. the same number of valence electrons as one boron and one nitrogen atom
together. The structure with space group 14 consists of layers - each of
them consisting of rings with four or eight atoms. 
Finally, two structures
with relatively large channels (i.e. large regions in the unit cell
without atoms) were found, with space groups 8 and
9, respectively.

The geometries are in reasonable agreement with the available experimental
data in table \ref{expstructures}.
The computed cell parameter $a$ and the
interlayer distance are approximatively constant for the layered
structures with only six-membered rings (with space groups 194, 160, 187);
this is also observed in
the experiment when comparing the $B_k$ and the rhombohedral structure.

Total energies and statistics are given in table \ref{StructuresEnergies}.
The five energetically lowest lying 
structures found were the layered structures
(space group 194, 160, 187) and the wurtzite and zincblende structure.
At zero pressure, LDA favors wurtzite and zincblende (by 0.05 $E_h$ per
four formula units), 
whereas B3LYP (by 0.01 $E_h$)
and Hartree-Fock favor the layered structures. This is in reasonable
agreement with other calculations, e.g. reference \cite{Albe1997}
and references therein
gives the layered structure by $\sim$0.06 eV/atom (0.02 $E_h$ per
four formula units) higher than the wurtzite and zincblende structures. 
Also, the zincblende and wurtzite structure
are nearly degenerate, and similarly the hexagonal and rhombohedral
structure. These results are stable with respect to the choice of the
basis set, as calculations with larger basis sets show (see appendix
\ref{basissetextrapolation}).

The statistics shows that the layered structures are frequently found,
as well as the 3d-structures such as wurtzite or zincblende. The statistics
includes all runs where Hartree-Fock energies were used during the
simulated annealing. 10 runs were performed, where the B3LYP functional 
was used during the simulated annealing procedure.
In two of these runs, a good structure candidate was found
(wurtzite and the layered
structure with space group 160). However, as was mentioned, the B3LYP
runs require much more CPU time, and therefore the runs were mainly
performed using the Hartree-Fock approach during the simulated annealing.
In total, about 16 \% of the runs gave one of the structure candidates
in table \ref{StructuresEnergies}. The other runs yielded either no
good structure candidates (like "amorphous" structures) or only energetically
very unfavorable structures.

When pressure is applied,
the lower coordinated structures become less favorable, which is in 
agreement with the rule that the coordination number increases with pressure,
see e.g.\cite{MuellerBuch}.
The enthalpies are displayed in figure \ref{LDAenthalpies} 
for LDA, and for B3LYP, respectively. Interestingly, in the case of
B3LYP the layered structures are
favorable at zero pressure, with wurtzite and zincblende becoming
favorable at a pressure of $\sim$ 3 GPa.

\clearpage

\begin{widetext}

\begin{figure}
\caption{ (Color online) 
The layered structures found, with space group 194, 160 and 187.
Green (light) spheres correspond to boron, blue (dark)
spheres to nitrogen atoms, 
respectively. The lines indicate the unit cells.}
\includegraphics[width=5cm]{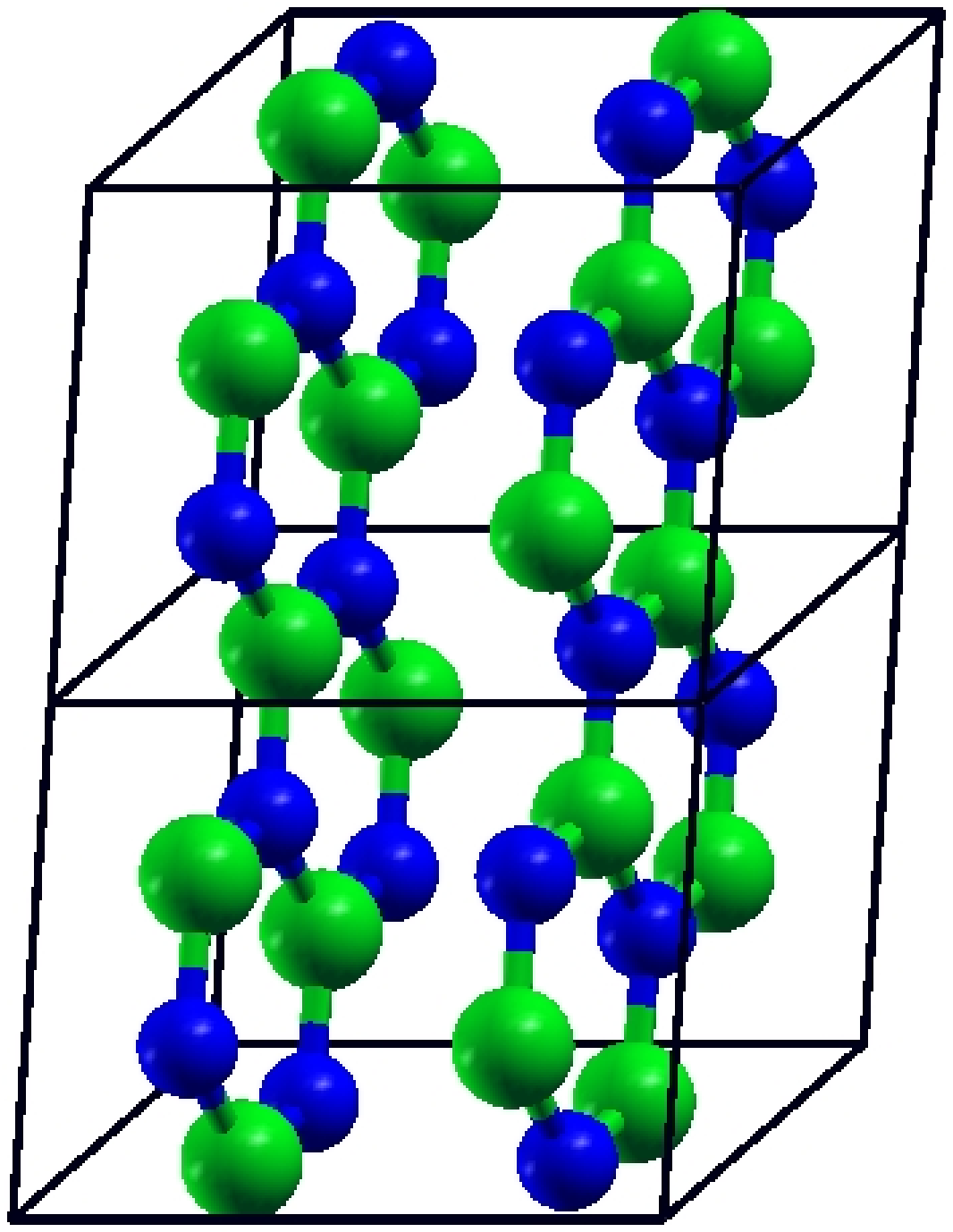} 
\includegraphics[width=5cm]{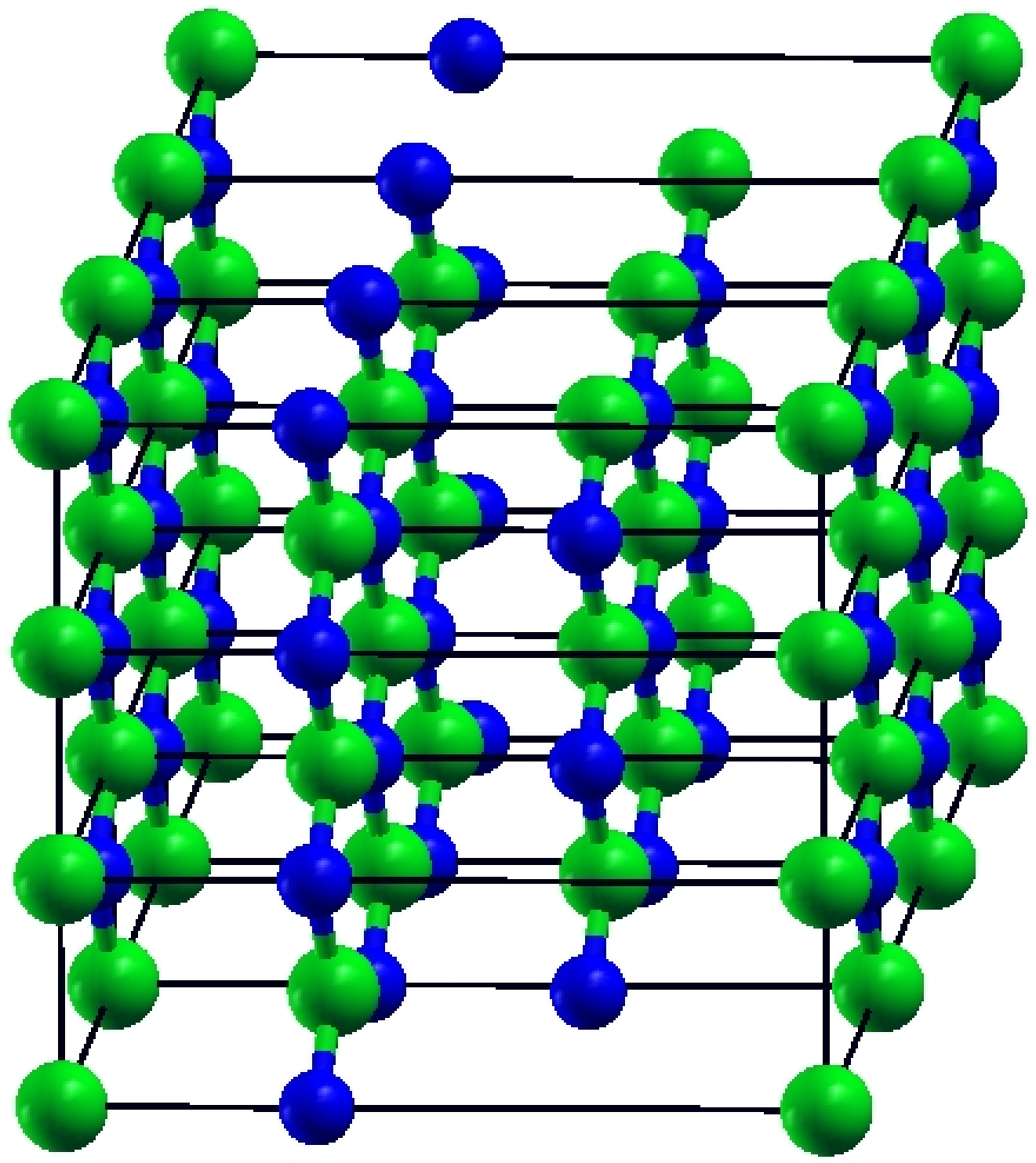} 
\includegraphics[width=5cm]{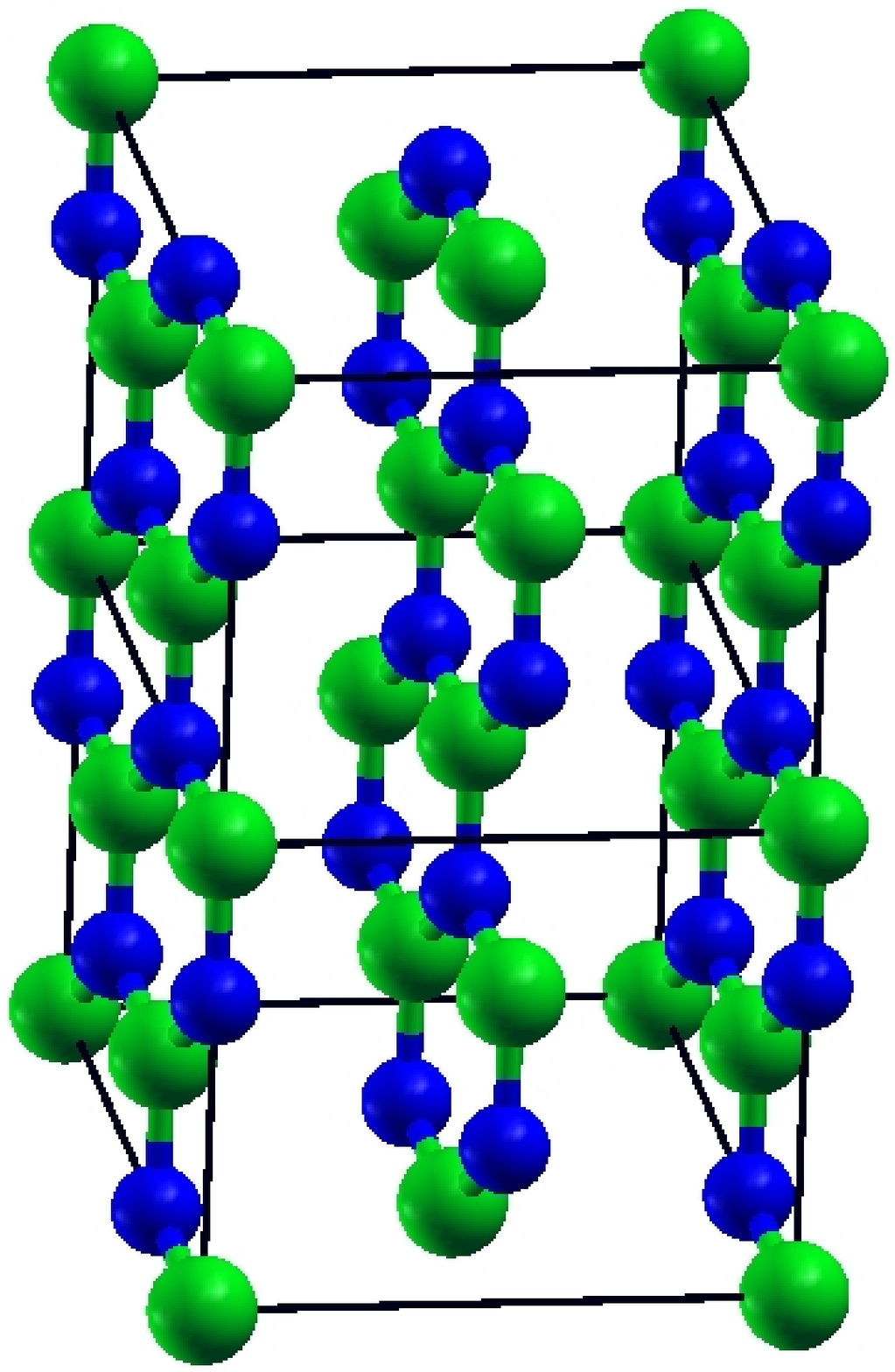} 
\label{schichten}
\end{figure}

\begin{figure}
\caption{ (Color online)  
The structures found, with space group 186, 216 and 136.}
\includegraphics[width=5cm]{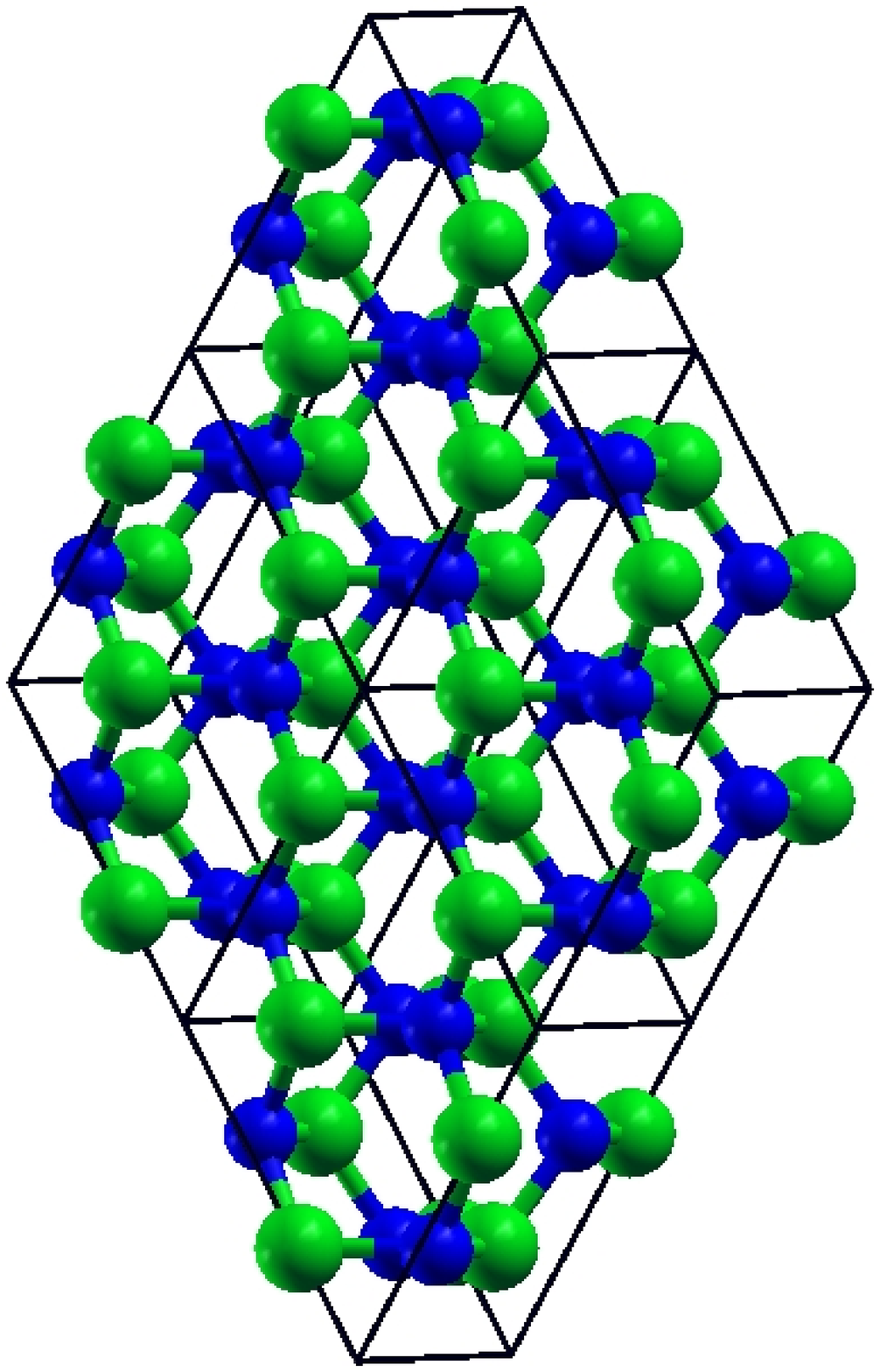} 
\includegraphics[width=5cm]{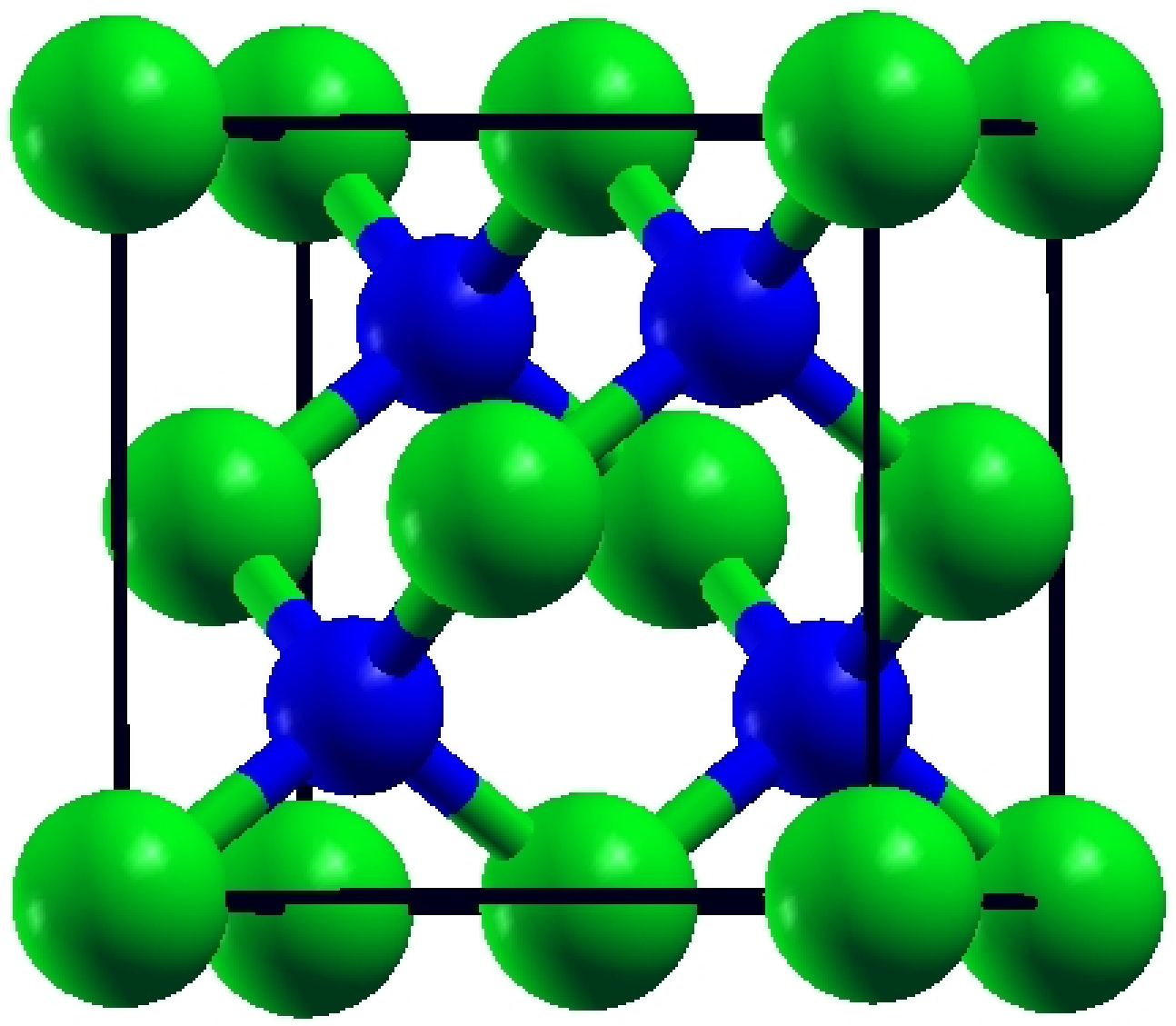} 
\includegraphics[width=5cm]{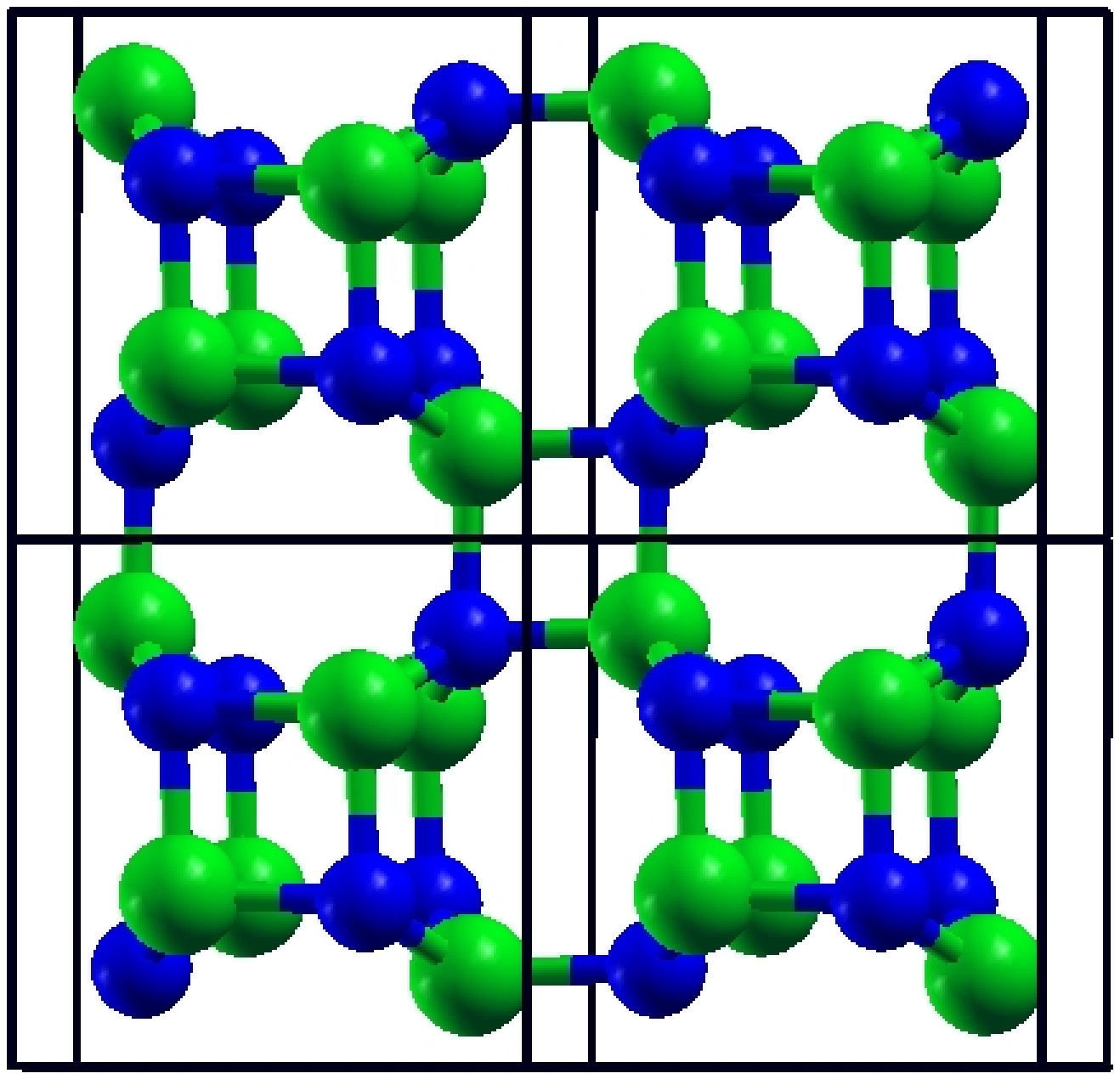} 
\label{186-216-136}
\end{figure}

\begin{figure}
\caption{ (Color online) The structures found, with space group 62 and 8.}
\includegraphics[width=5cm]{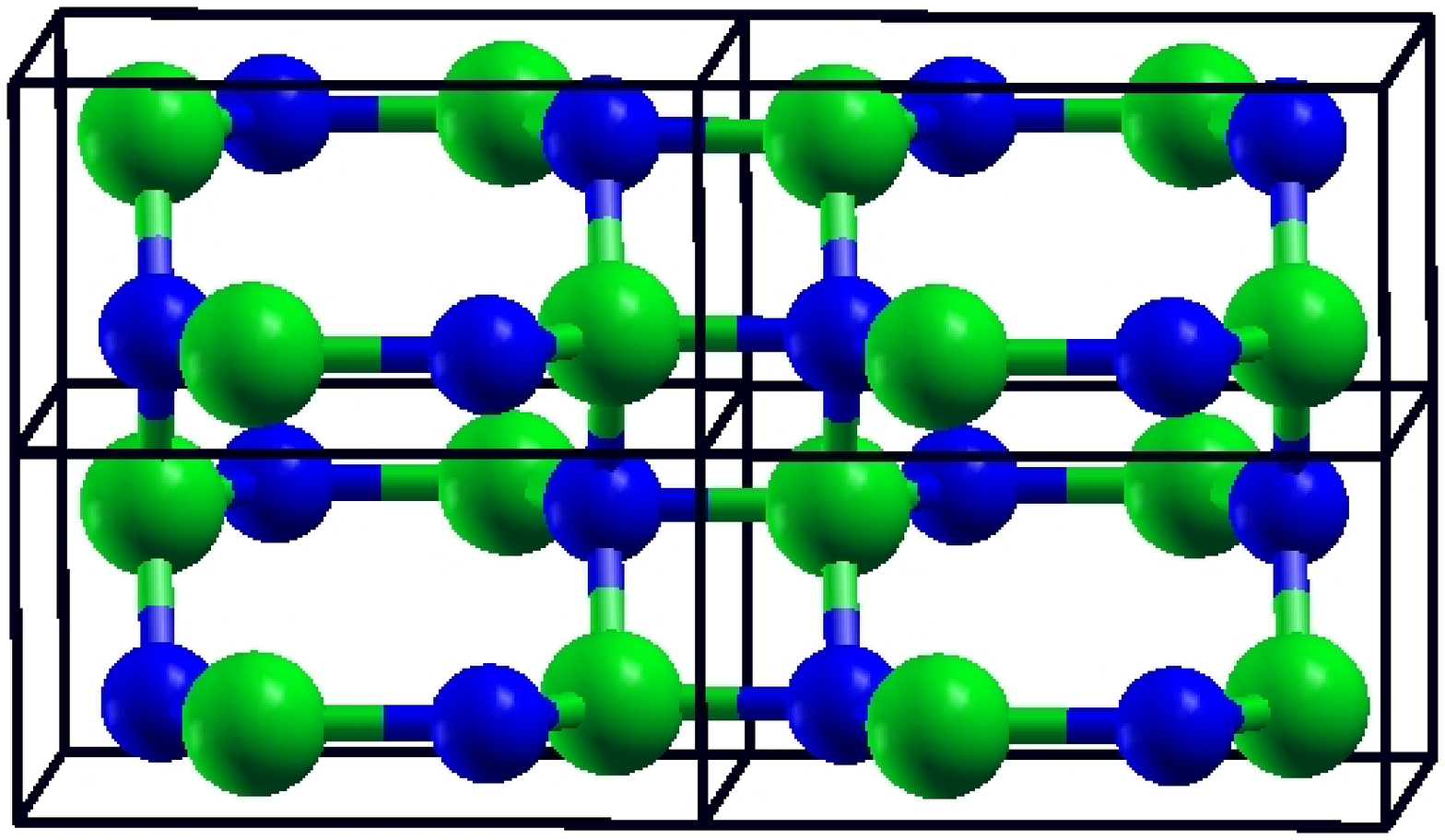} 
\includegraphics[width=5cm]{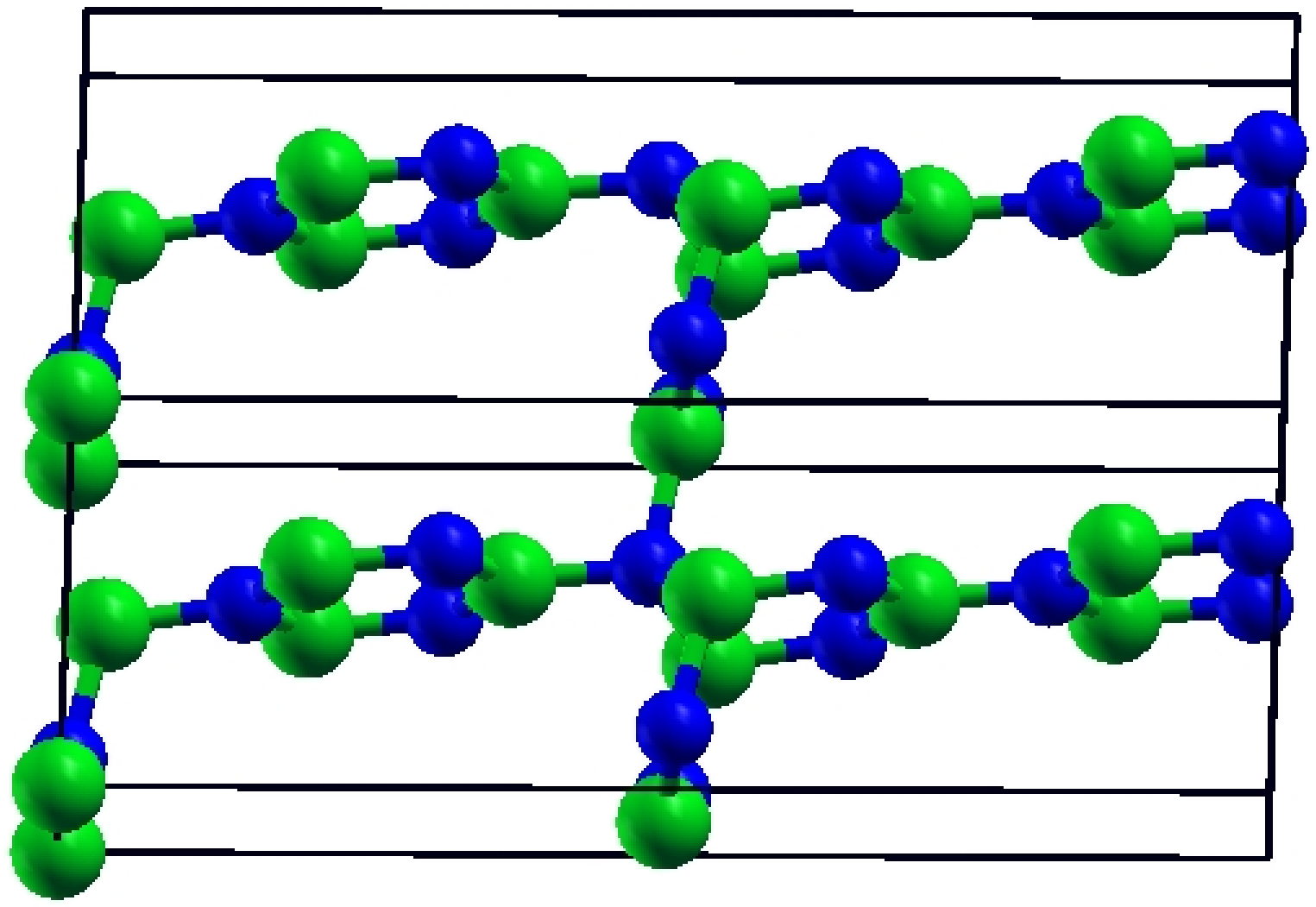} 
\label{62-8}
\end{figure}
\clearpage

\begin{figure}
\caption{(Color online)  The structures found, with space group 14 and 9.}
\includegraphics[width=5cm]{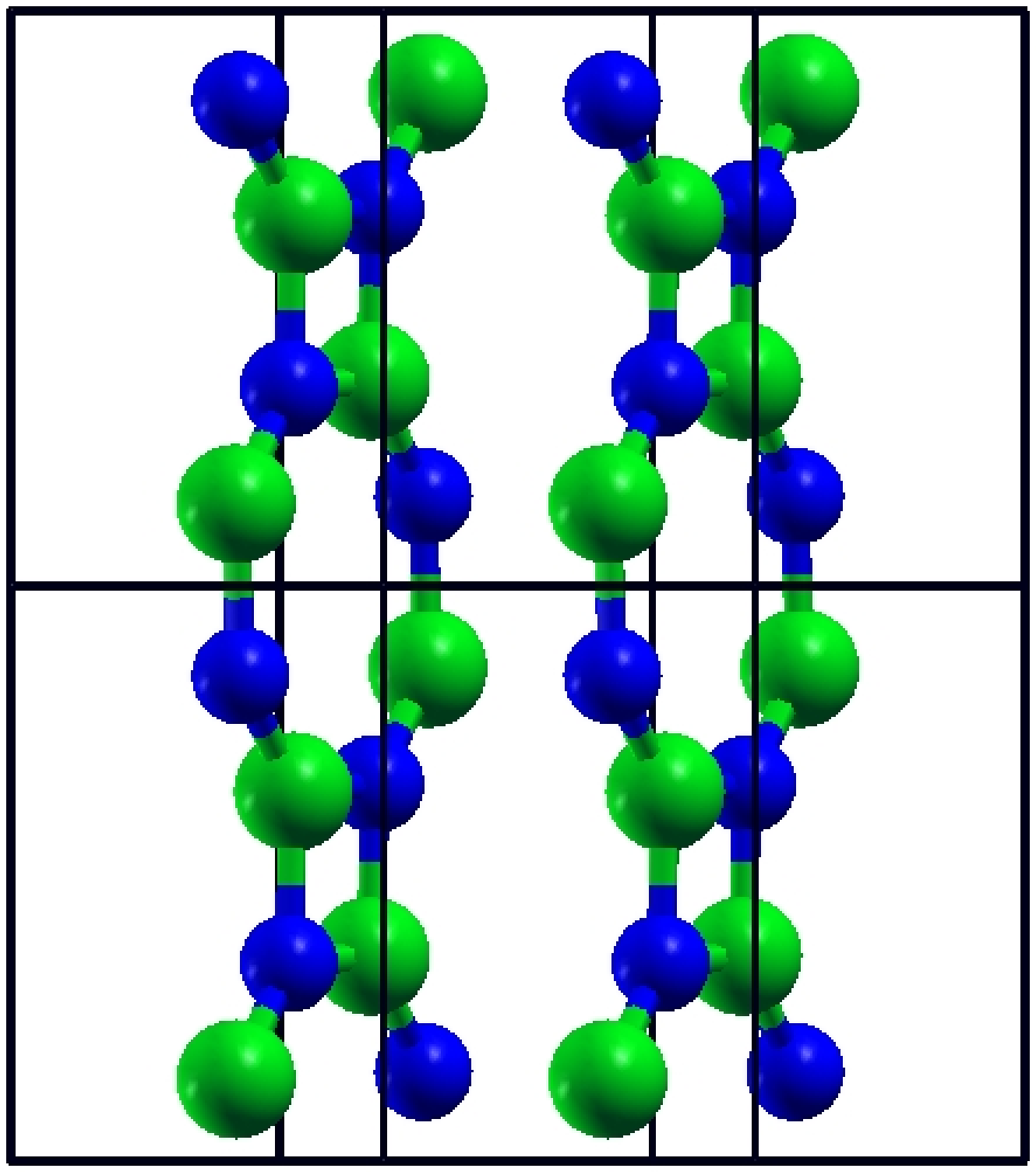} 
\includegraphics[width=5cm]{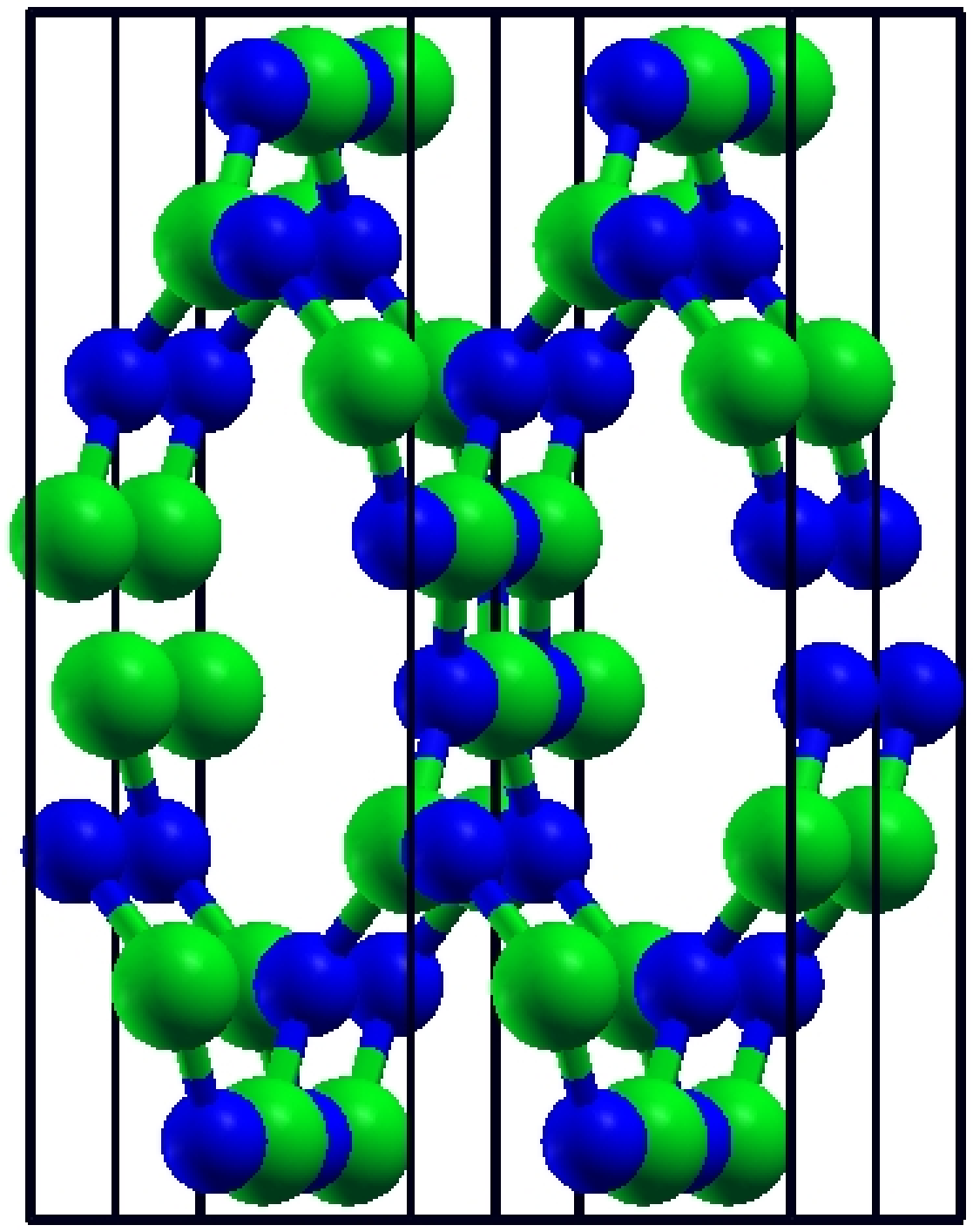} 
\label{14-9}
\end{figure}

\begin{figure}
\caption{(Color online) 
The enthalpy of the most relevant structures, at the LDA and
B3LYP level.
Structure candidates are labeled by their space group.}
\hspace{-2cm}
\includegraphics[width=7.5cm,angle=270]{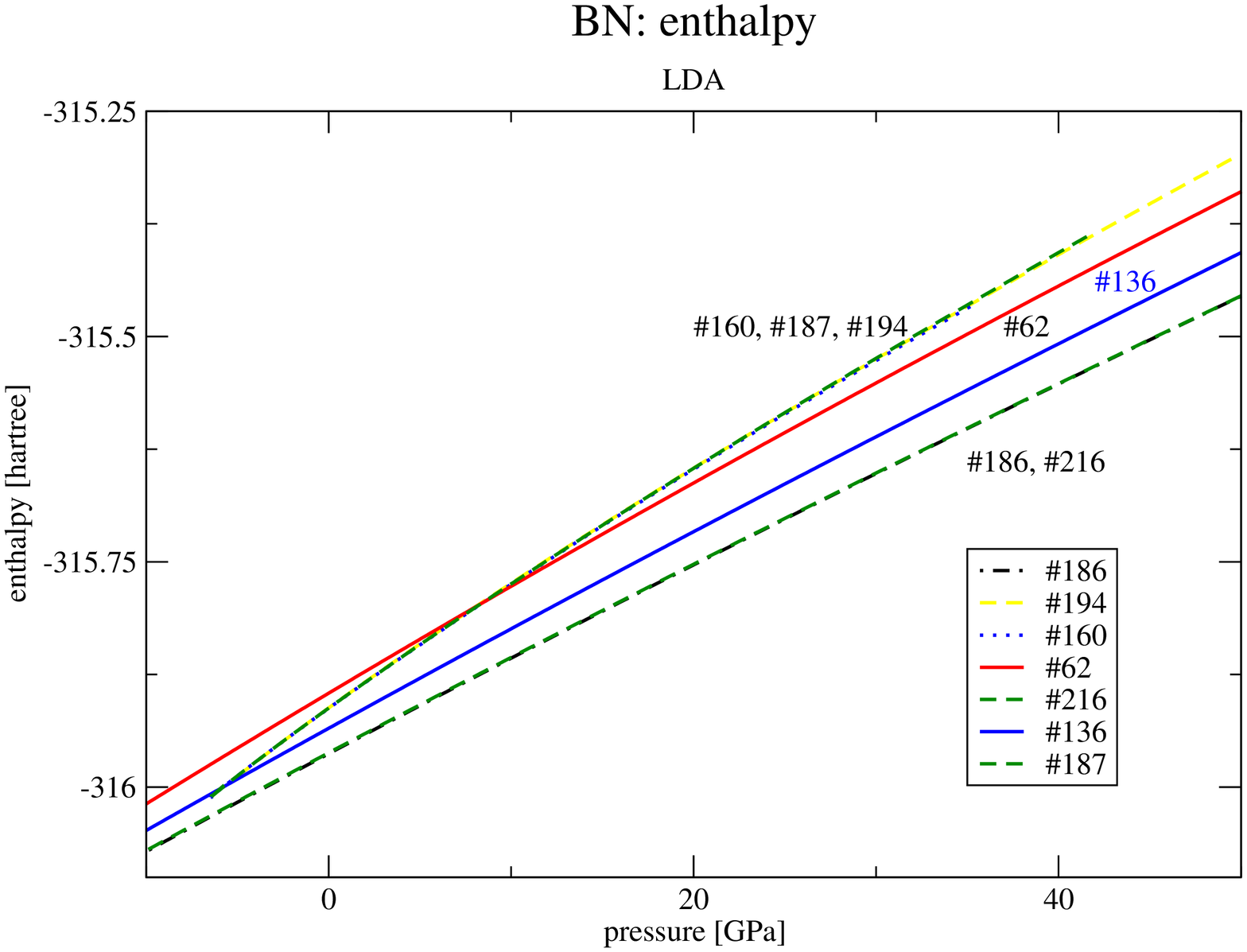}
\hspace{-2cm}
\includegraphics[width=7.5cm,angle=270]{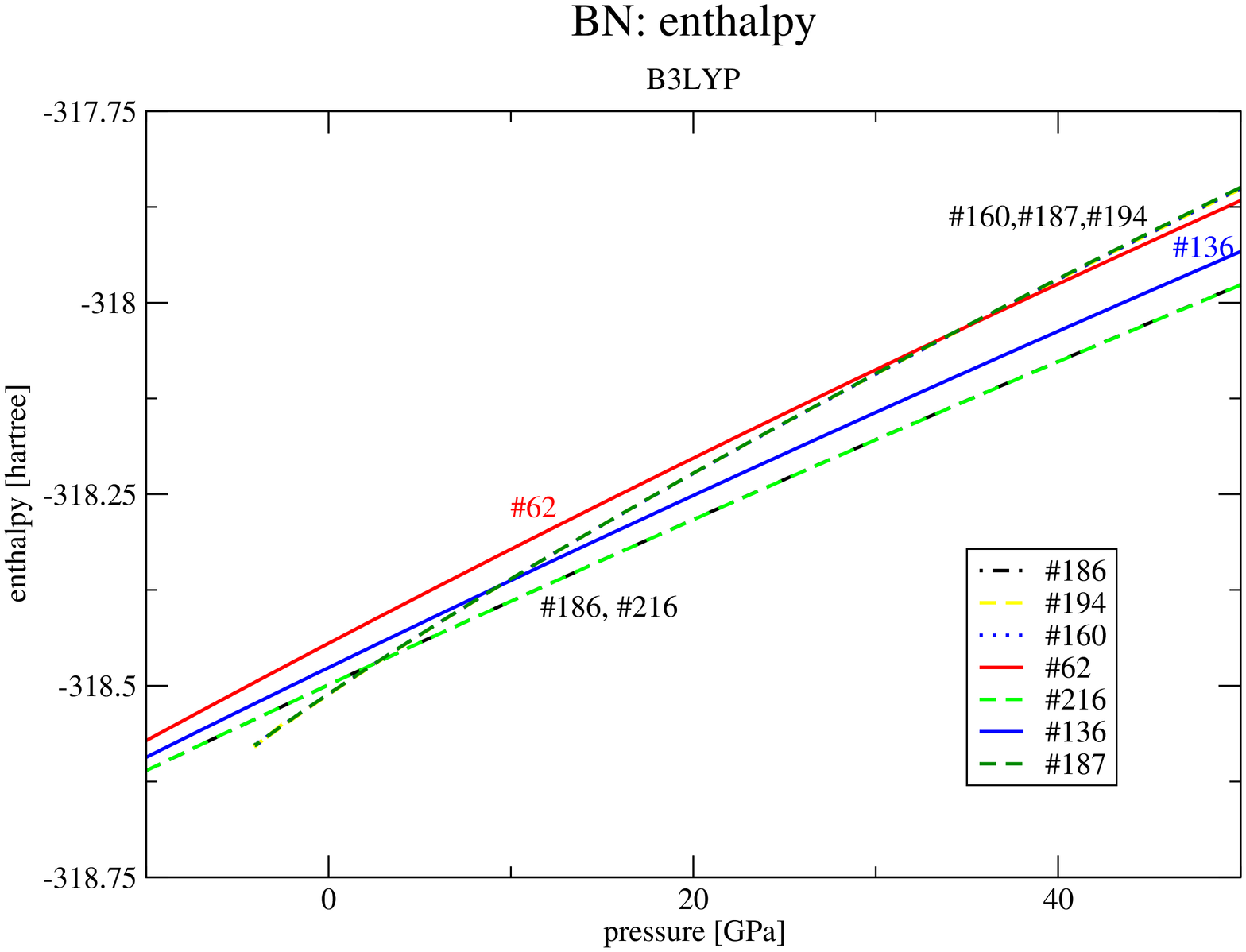} 
\hspace{-2cm}
\label{LDAenthalpies}
\end{figure}
\clearpage

\end{widetext}

\begin{widetext}

\begin{table}
\begin{center}
\caption{\label{Basissetssimannlocopt}
Basis sets used for the global search (I, II) and the local optimization
(III)}
\vspace{5mm}
\begin{tabular}{cccccc}
\hline\hline
\multicolumn{2}{c}{basis set I} & \multicolumn{2}{c}
{basis set II} &
\multicolumn{2}{c}{basis set III} \\ 
exponent & contraction &  exponent & contraction & exponent & contraction \\ 
\hline \multicolumn{6}{c}{B} \\
        \multicolumn{6}{c}{$s$}  \\ 
   2.082E+03& 1.850E-03 &    2.082E+03& 1.850E-03 &    2.082E+03& 1.850E-03  \\
   3.123E+02& 1.413E-02 &    3.123E+02& 1.413E-02 &    3.123E+02& 1.413E-02  \\
   7.089E+01& 6.927E-02 &    7.089E+01& 6.927E-02 &    7.089E+01& 6.927E-02  \\
   1.985E+01& 2.324E-01 &    1.985E+01& 2.324E-01 &    1.985E+01& 2.324E-01  \\
   6.292E+00& 4.702E-01 &    6.292E+00& 4.702E-01 &    6.292E+00& 4.702E-01  \\
   2.129E+00& 3.603E-01 &    2.129E+00& 3.603E-01 &    2.129E+00& 3.603E-01  \\
         \multicolumn{6}{c}{$sp$}  \\
   2.282E+00 &-3.687E-01 2.312E-01 \ &   2.282E+00 & -3.687E-01 2.312E-01 \
 & 2.282E+00 &-3.687E-01 2.312E-01 \\
 - &  & 4.652E-01 & 1.199E+00 8.668E-01 \ &  4.652E-01& 1.199E+00 8.668E-01 \\
 \multicolumn{6}{c}{$sp$} \\
0.4 & 1.0 1.0 & 0.25 & 1.0 1.0 & 0.197 & 1.0 1.0 \\
\multicolumn{6}{c}{$d$} \\
- &  & - & & 0.8 & 1.0\\
\multicolumn{6}{c}{N}\\
\multicolumn{6}{c}{$s$}\\
 4150.0   & 0.001845 &  4150.0   & 0.001845 &  4150.0   & 0.001845  \\
  620.1   & 0.01416 &    620.1   & 0.01416 &    620.1   & 0.01416   \\
  141.7   & 0.06863 &    141.7   & 0.06863 &    141.7   & 0.06863   \\
   40.34  & 0.2286 &      40.34  & 0.2286 &      40.34  & 0.2286    \\
   13.03  & 0.4662 &      13.03  & 0.4662 &      13.03  & 0.4662    \\
    4.47  & 0.3657 &       4.47  & 0.3657 &       4.47  & 0.3657    \\
\multicolumn{6}{c}{$sp$} \\
    5.425 & -0.4133 0.238 &   5.425 & -0.4133 0.238 & 5.425 & -0.4133 0.238 \\
    1.149 & 1.224  0.859 &    1.149 & 1.224  0.859 &  1.149 & 1.224  0.859 \\
\multicolumn{6}{c}{$sp$} \\
    0.3 & 1.0  1.0 & 0.297 & 1.0    1.0 & 0.297 & 1.0    1.0 \\
\multicolumn{6}{c}{$d$} \\
- &  & - & & 0.8 & 1.0\\ \hline\hline
\end{tabular}
\end{center}
\end{table}
\clearpage


\begin{table}
\begin{center}
\caption{\label{StructuresEnergies}Total energies of the most relevant
structures, and statistics. Energies
are in hartree units (1 $E_h$=27.2114 eV), for 4 formula units. 
A run is considered successful,
if one of the most relevant structures, as displayed in this table, was found.}
\vspace{5mm}
\begin{tabular}{ccccccc}
\hline\hline
name of modification & space group & 
\multicolumn{3}{c}{energy $[E_h]$} 
& \multicolumn{2}{c}{number of times found} \\
   &   & LDA & B3LYP & HF & basis I &  basis II \\ \hline
hexagonal BN & 194 &   -315.9121 &-318.5115 & -316.8753 & 1 & 2 \\
I-BN & 160 &   -315.9125 & -318.5110 & -316.8740 & 11 & 4 \\
II-BN & 187 &   -315.9123 & -318.5109 & -316.8739 & 0 & 2\\
wurtzite & 186 &   -315.9629 & -318.4988 & -316.8452 & 8 & 2 \\
zincblende & 216 &   -315.9619 & -318.4991 & -316.8459 & 4 & 2\\
$\beta$-BeO & 136 &   -315.9347 & -318.4753 & -316.8147 & 2 & 1\\
III-BN &  62 &   -315.8958 & -318.4437 & -316.7821 & 4 & 2\\
IV-BN &   8 &   -315.8810 & -318.4685 & -316.8243 & 6 & 0\\
V-BN &   9 &   -315.8707 & -318.4569 & -316.8138 & 2 & 0\\
VI-BN &  14 &   -315.8075 & -318.4085 & -316.7594 & 0 & 1 \\
\\
number of successful runs & & & & & 36 (18.9 \%) & 16 (11.5 \%) \\
number of runs in total & & & & & 190 & 139 \\ \hline\hline
\end{tabular}           
\end{center}
\end{table}


\begin{table}
\begin{center}
\caption{\label{Structuresfound}The energetically most favorable structures
found.}
\vspace{5mm}
\begin{tabular}{cccccc}
\hline\hline 
space group and 
&  \multicolumn{3}{c}{cell parameters and fractional coordinates}  \\
modification   &  LDA & B3LYP & HF  \\  \hline
194 & a=2.50 \AA, c=5.88 \AA & a=2.51 \AA, c=6.40 \AA 
& a=2.50 \AA, c=6.43 \AA \\
$B_k$
    & B (1/3, 2/3, 1/4) & B (1/3, 2/3, 1/4)  & B (1/3, 2/3, 1/4)\\
hexagonal BN & N (1/3, 2/3, 3/4) & N (1/3, 2/3, 3/4)  & N (1/3, 2/3, 3/4) \\ \\
160 & a=2.50 \AA, c=8.72 \AA & a= 2.51 \AA, c=9.56 \AA & a=2.50 \AA, 
      c=9.69 \AA \\
I-BN    & B (0, 0, 0) & B (0, 0, 0) & B (0, 0, 0)\\
    & N (1/3, 2/3, -0.0009) & N (1/3, 2/3, -0.0002) & N (1/3, 2/3, 0.) \\ \\
187 & a=2.50 \AA, c=5.83 \AA & a=2.51 \AA, c=6.38 \AA & a=2.50 \AA,  
      c=6.47 \AA \\
II-BN    & B (0, 0, 0) & B (0, 0, 0) & B (0, 0, 0) \\
    & B (1/3, 2/3, 1/2) & B (1/3, 2/3, 1/2) & B (1/3, 2/3, 1/2) \\
    & N (2/3, 1/3, 1/2) & N (2/3, 1/3, 1/2) & N (2/3, 1/3, 1/2) \\
    & N (1/3, 2/3, 0) & N (1/3, 2/3, 0) & N (1/3, 2/3, 0)\\ \\
186 & a=2.54 \AA, c=4.19 \AA & a=2.57 \AA, c=4.23 \AA & a=2.55 \AA,
      c=4.21 \AA \\
wurtzite 
    & B (2/3, 1/3, 0) & B (2/3, 1/3, 0) & B (2/3, 1/3, 0) \\
    & N (2/3, 1/3, 0.3748) & (2/3, 1/3, 0.3750) & (2/3, 1/3, 0.3752)\\ \\
216 & a=3.60 \AA & a=3.64 \AA & a=3.62 \AA  \\
zincblende 
    & B (0, 0, 0) & B (0, 0, 0) & B (0, 0, 0) \\
    & N (1/4, 1/4, 1/4) & N (1/4, 1/4, 1/4) & N (1/4, 1/4, 1/4)\\ \\
136 & a=4.38 \AA, c=2.54 \AA & a=4.43 \AA, c=2.56 \AA & a=4.41 \AA,
      c=2.55 \AA \\
$\beta$-BeO    & B (-0.1738, 0.1738, 1/2) & B (-0.1744, 0.1744, 1/2) & 
(-0.1742, 0.1742, 1/2)\\
    &  N (-0.1880, -0.1880, 1/2) & N (-0.1872, -0.1872, 1/2) & 
(-0.1872, -0.1872, 1/2)\\ \\
 62 & a=4.76 \AA, b=2.58 \AA, c=4.29 \AA & a=4.85 \AA, b=2.60 \AA,
      c=4.31 \AA &
a=4.83 \AA, b=2.59 \AA,   c=4.30 \AA \\
III-BN    & B (-0.3404, 3/4, 0.0921) & B (-0.3397, 3/4, 0.0912) & 
B (-0.3399, 3/4, 0.0905)\\
    & N (0.3171, 3/4, 0.1048) & N (0.3211, 3/4, 0.1074)& 
N (0.3215, 3/4, 0.1078)\\ \\
  8 & a=12.95 \AA, b=2.50 \AA, c=4.33 \AA  & 
      a=13.08 \AA, b=2.51 \AA, c=4.39 \AA  & 
      a=13.03 \AA, b=2.50 \AA, c=4.37 \AA \\
IV-BN    & $\beta$=91.7$^\circ$ & $\beta$=90.7$^\circ$ & $\beta$=90.7$^\circ$ \\
    & B  (0, 0, 0) & B  (0.0001, 0, -0.0011) & B  (0.0002, 0, -0.0008)   \\
    &  B  (0.1998, 0., 0.4266) & B (0.2000, 0., 0.4268)
 & B  (0.2003, 0., 0.4265) \\
    &  B  (-0.1342, 0., 0.3784) & B (-0.1344, 0., 0.3806) 
& B  (-0.1343, 0., 0.3820)\\
    &  B  (0.0319, 1/2, -0.4995) & B (0.0319, 1/2, 0.4995)
& B (0.0320, 1/2, 0.4998) \\
    &  N  (-0.0192,  0.,  0.3442) & N (-0.0192, 0., 0.3436)
& N (-0.0194, 0., 0.3430)\\
    &  N  (0.0062, 1/2, -0.1562) & N (0.0057, 1/2, -0.1562)
& N (0.0051, 1/2, -0.1563)\\
    &  N  (0.1468, 1/2, 0.4443) & N (0.1471, 1/2, 0.4424)
& N (0.1471, 1/2, 0.4409)\\
    &  N  (-0.1878,  1/2,  0.3908) & N (-0.1876, 1/2, 0.3928)
& N (-0.1875, 1/2, 0.3934) \\ \\
  9 & a=3.00 \AA, b=9.48 \AA, c=4.33 \AA
    & a=3.21 \AA, b=9.46 \AA, c=4.35 \AA
    & a=3.24 \AA, b=9.42 \AA, c=4.33 \AA \\
V-BN & $\beta$=105.1$^\circ$ & $\beta$=105.8$^\circ$ &  $\beta$= 106.0$^\circ$ \\
    & B (-0.4996, -0.0652, 0.0001) & B (-0.4991, -0.0657, 0.0004) 
& B (-0.4985, -0.0658, 0.0006) \\
    &  B (0.3780, -0.1931, 0.4920) & B (0.3717, -0.1931, 0.4920) 
& B (0.3711, -0.1929, 0.4923)\\
    &  N (-0.4219, -0.1948, -0.1601)& N (-0.4173, -0.1955, -0.1593)
& N (-0.4160, -0.1953, -0.1592)\\
    &  N (0.4422, -0.0651, 0.3218)& N (0.4435, -0.0654, 0.3209)
& N (0.4423, -0.0653, 0.3202)\\ \\
 14 & a=3.17 \AA,  b=4.90 \AA, c=4.93 & a=3.44 \AA, b=4.92
 \AA, c=4.94 \AA, 
 & a=3.48 \AA,  b=4.90 \AA, c=4.91 \AA \\
VI-BN & $\beta$=117.7$^\circ$ & $\beta$=115.1$^\circ$ & $\beta$=114.3$^\circ$ \\
    & B (0.4963, -0.1429, 0.1385) & B (-0.4998, -0.1428, 0.1413) &
B (-0.4992, -0.1427, 0.1417) \\
    & N (0.4984, 0.3443 -0.3418) & N (0.4996, 0.3437, -0.3427) &
N (0.4989, 0.3441, -0.3437) \\ \hline\hline
\end{tabular}           
\end{center}
\end{table}
\end{widetext}

\clearpage

\section{Conclusion}
It was shown that structure prediction based on simulated annealing
and using {\em ab-initio} energies during both the global and local
optimization is feasible for a covalent system such as boron nitride. This is
a significant extension of the previous work \cite{DollPCCP}
where this approach
was shown to be feasible for an ionic system. Covalent systems are
more difficult to study as covalent bonds need to be established between
the neighbors, and convergence problems are more severe in this case.
Three layered structures, the wurtzite and zincblende structure, 
a structure of the $\beta$-BeO type, and four other
favorable structures were found. Applying
pressure leads to a preference of the higher coordinated structures.

\appendix

\section{Larger basis sets to extrapolate the basis set limit}
\label{basissetextrapolation}
In the present work, a main task is to compute the
energy differences between various structures. Besides the functional,
the choice of the basis set has an influence on these
results. In order to investigate this in more detail, the basis set
used for the local optimization was further enlarged, and the total
energies were computed for the most important structures. Here, enlarging
the basis set means to include more diffuse functions. It turned
out that this was only possible for the nitrogen atom, whereas more
diffuse functions on boron led to linear dependence problems.
Therefore, in a first step, one $sp$ shell with exponent 0.15 was
added to the nitrogen basis set III in table \ref{Basissetssimannlocopt},
which resulted in a $[4s3p1d]$ basis set for nitrogen (basis set IV).
In a second step, two $sp$ shells (with exponents 0.15 and 0.6) were
added to the nitrogen basis, i.e. a $[5s4p1d]$ basis set was obtained
(basis set V). 
Note that these basis sets work reasonably well at
zero pressure, but numerical instability sets in with compression, i.e.
enthalpies (as in figure \ref{LDAenthalpies}) 
could only be obtained up to
a relatively small pressure.

\begin{figure}
\caption{Total energies for the various structures, in hartree per four
formula units, for basis sets
III, IV and V, at the geometry optimized for each basis set. Structure
candidates are labeled by the space group}
\vspace{2cm}
\includegraphics[width=8cm]{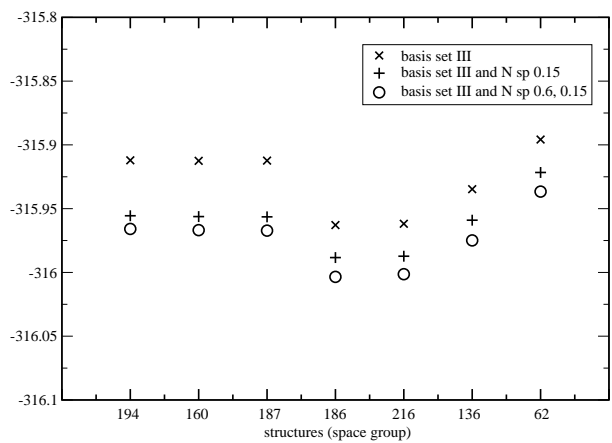} 
\label{basissatzvergleich}
\end{figure}

A geometry optimization was performed with these basis sets.
The results are displayed in table \ref{Basisbenchmark}.
It becomes obvious, that the energy differences between the various
structures remain essentially constant; the total energy
becomes lower with increasing basis set.
This is visualized in figure \ref{basissatzvergleich}. 

The geometry slightly changes when enlarging the basis set. Most prominent
is the change of the $c$-axis for the layered structures
when the basis set is enlarged. This is due to the weak bonding between
the individual layers. However, as a whole, the basis set does not
change the relative energies between the structures.

\begin{widetext}
\begin{table}
\begin{center}
\caption{\label{Basisbenchmark}A comparison of three basis sets for the
energetically most favorable structures found.}
\vspace{5mm}
\begin{tabular}{cccc} \hline\hline
 space group  &  basis set & 
\multicolumn{1}{c}{cell parameters, in \AA} & \ energy, in $E_h$ \ \\
and modification &  & (LDA)  \\ \hline
194 & III & a=2.50 c=5.88  & -315.9121 \\
hexagonal BN    &  IV & a=2.51 c=6.24  & -315.9556 \\
    &   V & a=2.51 c=6.18  & -315.9659 \\ \\

160 & III & a=2.50 c=8.72  & -315.9125 \\
I-BN    &  IV & a=2.51 c=9.27  & -315.9562 \\
    &   V & a=2.50 c=9.19 & -315.9668 \\ \\

187 & III & a=2.50 c=5.83  & -315.9123 \\
II-BN    &  IV & a=2.51 c=6.18 & -315.9564 \\
    &   V & a=2.50 c=6.12 & -315.9672 \\ \\

186 & III & a=2.54 c=4.19  & -315.9629\\
wurtzite    &  IV & a=2.54 c=4.19  & -315.9884 \\   
    &   V & a=2.54 c=4.18 & -316.0035 \\ \\

216 & III & a=3.60 & -315.9619\\
zincblende    &  IV & a=3.60 & -315.9873 \\
    &   V & a=3.60 & -316.0014  \\ \\

136 & III & a=4.38 c=2.54 & -315.9347\\
III-BN    &  IV & a=4.39 c=2.54 & -315.9591 \\
    &   V & a=4.37 c=2.54 & -315.9749 \\ \\

 62   & III & a=4.76 b=2.58 c=4.29 & -315.8958\\
IV-BN &  IV & a=4.79 b=2.58 c=4.29 & -315.9216\\
      &   V & a=4.76 b=2.57 c=4.29 & -315.9366 \\
\hline\hline
\end{tabular}           
\end{center}
\end{table}
\end{widetext}

\clearpage
\newpage

\clearpage

\acknowledgments
We would like to thank Prof. D. Proserpio (Milano) for a valuable
discussion.

\end{document}